\RequirePackage[T1]{fontenc}
\RequirePackage[utf8]{inputenc}

\documentclass[%
 reprint,
superscriptaddress,
 amsmath,amssymb,
 aps,
]{revtex4-2}
\usepackage[utf8]{inputenc}
\usepackage{newunicodechar}
\newunicodechar{─}{--}

\usepackage{graphicx}
\usepackage{dcolumn}
\usepackage{bm}
\usepackage{hyperref}
\usepackage[mathlines]{lineno}
\usepackage{xcolor}
\usepackage{newunicodechar}
\newunicodechar{ₓ}{$_x$}
\usepackage[version=4]{mhchem}
\newcommand{\rev}[1]{\textcolor{black}{#1}}
\usepackage{makecell}
\usepackage{comment}
\usepackage{hyperref}

\begin{document}

\title{\textit{Ab initio} Monte Carlo prediction of order-to-disorder transition\rev{s} in multicomponent MXenes}

\author{Noah Oyeniran}
\affiliation{Department of Aerospace Engineering and Mechanics, The University of Alabama, Tuscaloosa, AL 35487, USA}
%
%
\author{Chongze Hu}
\email{hucz@ua.edu}
\affiliation{Department of Aerospace Engineering and Mechanics, The University of Alabama, Tuscaloosa, AL 35487, USA}

\date{\today}

\begin{abstract}
This letter \rev{predicts previously unreported order-to-disorder transition behaviors in multicomponent MXenes using an integrated and improved} first-principles Monte Carlo (MC) framework. 
\rev{The improvements include ($i$) structural relaxation and ($ii$) selective atom swapping during MC iterations for more accurate and efficient predictions}. 
Using (TiMo)-based double transition metal (DTM) carbide MXenes as a model system, \textit{ab initio} MC simulations reveal that \rev{surface termination and coordination environments play critical roles in governing chemical ordering in MXenes. 
Specifically,}
the formation of out-of-plane MXene (o-MXenes) with Mo segregation to outermost metallic layers (M') is only driven by the oxygen (O) termination at prismatic sites. 
In contrast, O termination at octahedral sites and fluorine (F) termination at both prismatic and octahedral sites always promote the formation of o-MXenes with Ti-segregated to M' layers. 
Furthermore, changing the F/O ratio at prismatic termination sites or alternating the atomic coordination within the MXene lattices can induce an order-to-disorder transition in DTM MXenes.    
%
%
\end{abstract}

\maketitle

\par{
MXenes are two-dimensional (2D) transition metal carbides and/or nitrides that represent an emerging class of low-dimensional materials with exceptional properties and performances~\cite{gogotsi2023future, naguib2012two, naguib2011am, anasori2017review, kamy2020science, li2022review, lim2022review, wyatt20212d}.
%
%
Their chemical formula is typically written as M$_{n+1}$X$_n$T$_x$, where M represents transition metal, X is nonmetallic element such as C or N atoms, T$_x$ is the surface termination (e.g., F, O, Cl, -OH, and others)~\cite{gogotsi2023future, naguib2012two, naguib2011am, anasori2017review, kamy2020science, li2022review, lim2022review, wyatt20212d}.
The large variety of possible M elements and surface terminations provides a vast compositional space of MXenes and highly tunable properties~\cite{thangavelu2025review, zhou2023atomic, rems2024cm}. 
Over the past decade, more than 50 MXenes have been experimentally synthesized and with over 100 additional MXenes being theoretically predicted~\cite{oyeniran2025afm, khan2023review, rems2024cm}.   

Inspired by design strategies used in high-entropy (HE) materials~\cite{oses2020natrevmater, han2024natrevmater, george2019natrevmater, miracle2017acta, qin2020jecs, qin2020scripta}, recent efforts have expanded MXenes into the HE regime by incorporating multiple M elements within a single lattice~\cite{hanan2025comprehensive, hussain2024double, chen2025high, das2023perspective, leong2022elucidating, zhang2025ultra, biswas2022mxene, nemani2021high, oyeniran2025apl}.
The presence of multiple transition metals greatly broadens the compositional space of MXenes, leading to substantial structural and chemical diversity~\cite{wyatt2025science, nemani2023commat, 2021science}.  
Notable examples of this diversity is the formation of chemically ordered MXenes in either in-plane or out-of-plane forms~\cite{hong2020mrs, 2021science}, which have been experimentally observed in a range of multicomponent MXenes, including double transition metals (DTM) Mxenes (e.g., Mo$_2$ScC$_2$T$_x$~\cite{meshkian2017acta}, Mo$_2$TiC$_2$T$_x$~\cite{anasori2015acs}, Cr$_2$TiC$_2$T$_x$~\cite{anasori2015acs}, Mo$_2$Ti$_2$C$_3$T$_x$~\cite{anasori2015acs}, and Ti$_2$TaC~\cite{maldonado2022acs}) and high-entropy MXenes featuring up to six metals~\cite{wyatt2025science}.
Owing to their superior properties and performance compared to conventional single transition metal MXenes~\cite{2021science}, tuning the chemical ordering is recognized as an important strategy for designing novel multicomponent MXenes. 
%

Despite this promising strategy, the effective design of chemically ordered MXenes is hindered by a poor understanding of the \rev{basic driving forces} that govern these ordering behaviors.  
\rev{Although configurational entropy is widely recognized as the dominant factor that induces order-to-disorder transitions in multicomponent material systems~\cite{2021science}, it remains unclear what additional factors contribute to these transition behaviors.
Using an improved density functional theory/Monte Carlo (DFT/MC) framework, we show that order-to-disorder transition behaviors can also be driven by surface termination and local coordination environments, and even occur in low-entropy, DTM MXenes.  
}  
This improved framework builds upon our prior work~\cite{oyeniran2025apl}, \rev{where new functionalities are introduced to account ($i$)} structural relaxation during each MC iteration and \rev{($ii$) selective atom swapping at any sites such as M, X, and T sites}. 
\rev{Within the scope of this work, swapping/mixing is only done at both M and T sites to better understand the order-to-disorder transitions in DTM MXenes}.
\rev{These improvements} allow the framework to function more accurately \rev{and efficiently} capture the local chemical ordering \rev{by identifying} the minimum energy configurations (MECs) of multicomponent MXenes. 
\begin{figure}[ht!]
\centering
\dimendef\prevdepth=0
{
\includegraphics[width=0.99\linewidth]{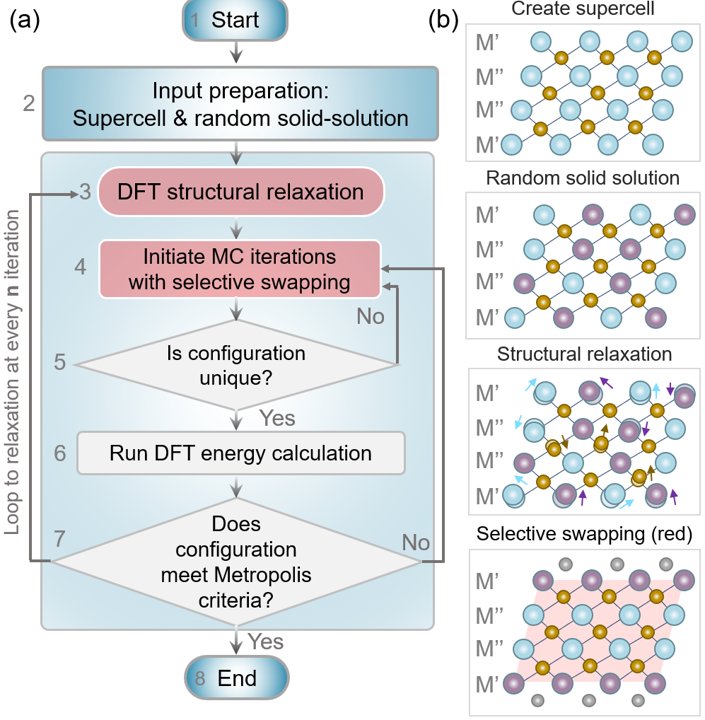}
}
\caption{
Integrated density functional theory (DFT) and Monte Carlo (MC) simulation framework to predict the minimum energy configurations (MECs) of multicomponent MXenes.
(a) Workflow of the DFT/MC framework \rev{with new functionalities include ($i$) structural relaxation and ($ii$) selective atom swapping} during the MC steps at certain interval. 
(b) Schematic diagrams showing the atomic configurations of multicomponent MXenes at different MC stages.
}
\label{fig:Fig1}
\end{figure}

}

\par
\begin{figure*}[ht!]
\centering
\dimendef\prevdepth=0
{
\includegraphics[width=0.95\textwidth]{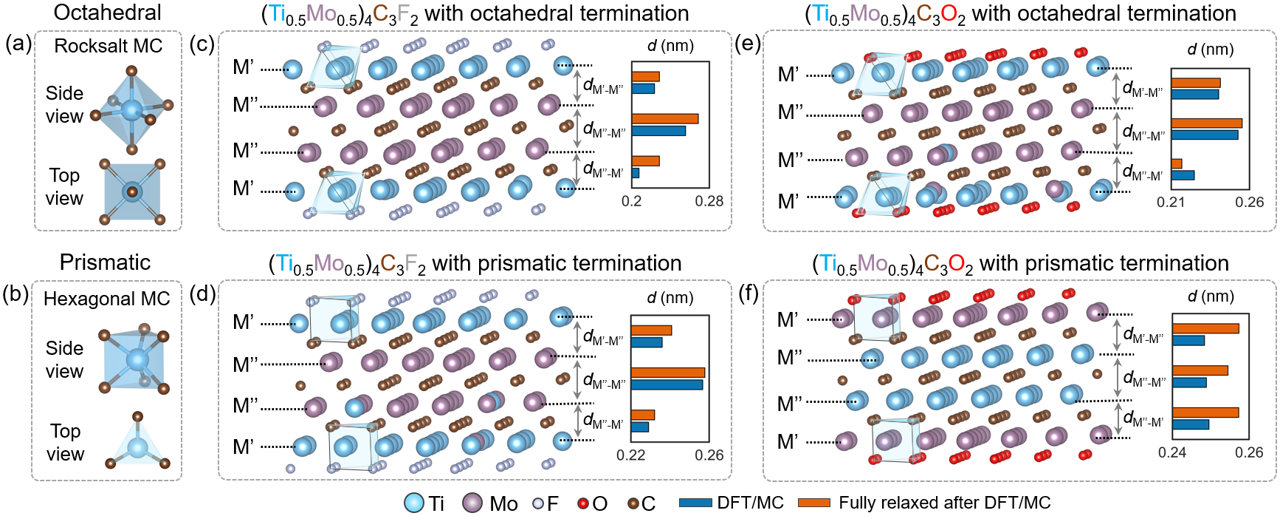}
}
\caption{Chemical ordering of (TiMo)-based DTM MXenes with different surface termination elements and coordination environments.  
(a-b) Schematics of octahedral and prismatic coordination in rocksalt and WC-hexagonal transition metal carbides (MC).
DFT/MC-identified MECs of (Ti\(_{0.5}\)Mo\(_{0.5}\))\(_4\)C\(_3\)F\(_2\) with (c) octahedral and (d) prismatic coordination of M' sites. 
DFT/MC-identified MECs of Ti\(_{0.5}\)Mo\(_{0.5}\))\(_4\)C\(_3\)O\(_2\) with (e) octahedral and (f) prismatic coordination of M' sites. 
For each panel, the interlayer distance ($d$) between M layers are determined based on DFT/MC-searched and fully relaxed structures. 
The energy profile $vs.$ MC iterations for each (TiMo)-based DTM MXene is shown in Fig. S1.
%
}
\label{fig:Fig2}
\end{figure*}

{
The improved DFT/MC workflow is illustrated schematically in~Fig.~\ref{fig:Fig1}(a). 
%
%
Similar to our prior work~\cite{oyeniran2025apl}, the workflow begins with the preparation of necessary input files and parameters, including a structural file that defines the initial configuration and the atomic fractions of substitution elements, \rev{as shown in}~Fig.~\ref{fig:Fig1}(b).
Based on these \rev{inputs}, the initial structures are constructed as supercells, \rev{and then specified} substitution elements are randomly inserted into the host matrix to form a solid-solution-like configuration \rev{(see~Fig.~\ref{fig:Fig1}(b))}.
The resulting supercell is then subjected to DFT structural relaxation to account for lattice distortions introduced by element substitution. 
%
The relaxed structure, along with its total energy and atomic fingerprint, is stored as the first \rev{MC} iteration.

Following initialization, the MC loop begins by generating \rev{a new atomic swapped configurational arrangement} through \rev{selective atom swapping. Here, only atoms in the specified regions or sites are subjected to swapping while others are fixed~\rev{(see~Fig.~\ref{fig:Fig1}(b))}}. 
%
%
For each new configuration, DFT energetic calculations are performed to obtain its ground-state energy.
Based on the energy differences between different configurations, the Metropolis criterion is applied to decide whether the new configuration is accepted or rejected~\cite{metropolis1953equation, oyeniran2025apl}, thus guiding the search toward energetically favorable states.
Once a stable configuration is identified during the MC loop, the system undergoes another DFT structural relaxation before proceeding to the next iteration. \rev{This structural relaxation is set to occur periodically at a certain number of iterations to save computational cost and ensure faster energy convergence.} 
\rev{In this study, about 5000 MC steps are performed} \rev{with relaxation taking place after every 30 iterations to drive to} fully converged energy profiles \rev{(i.e., energy reaching a plateau)}. 
\rev{The attainable configurations are ultimately considered the MECs, which are the most stable and synthesizable structural arrangements of the multicomponent MXenes under consideration}.
\rev{See Supplementary Notes 1-3 and Figs. S1-3 for full energy convergence landscapes of all MXenes studied in this work}. 
%
%

All DFT \rev{energetic} calculations were performed using the plane-wave Vienna \textit{Ab initio} Simulation Package (VASP)~\cite{kresse1993ab, kresse1996efficiency, kresse1996efficient}, employing the projector augmented-wave (PAW) method~\cite{blochl1994projector, kresse1999ultrasoft} and the generalized gradient approximation (GGA)} parameterized by Perdew–Burke–Ernzerhof (PBE)~\cite{perdew1996generalized} for the exchange–correlation functional. 
Van der Waals corrections were included using the PBE-D2 approach~\cite{grimme2006semiempirical}, which has been demonstrated to accurately capture interlayer interactions in layered materials such as MXenes~\cite{oyeniran2025apl}. 
The plane-wave cutoff energy was set to 500~eV, with energy and force convergence criteria of $10^{-3}$~eV and 0.01~eV~\AA$^{-1}$, respectively. 
The Brillouin zone was sampled using a \textit{$\Gamma$}-centered $k$-point mesh of $4 \times 4 \times 1$. 
\rev{Spin polarization is not considered due to the small magnetization in (TiMo)-based MXenes, see Supplementary Note 4 and Fig. S4.}
We first examine the effects of surface terminations, \rev{specifically F and O atoms,} and their coordination environments in equimolar (TiMo)-based DTM carbide MXenes. 
Two distinct coordination environments, octahedral and prismatic (Fig.~\ref{fig:Fig2}(a-b)), originating from rocksalt and hexagonal transition metal carbides~\cite{hu2018acs, hu2018jpcc}, were considered for surface termination.   
Figures~\ref{fig:Fig2}(c-d) illustrate the MECs of $(\mathrm{Ti}_{0.5}\mathrm{Mo}_{0.5})_{4}\mathrm{C}_{3}\mathrm{F}_{2}$ with F terminated at both octahedral and prismatic sites. 
These MECs consistently demonstrate the formation of o-MXenes, where Ti atoms segregate into the outermost M' layers while Mo atoms remain in the inner M'' layers, regardless of whether the surface coordination of F atoms is octahedral or prismatic. 
By calculating the energy difference, the octahedral coordination is lower in energy than the prismatic case by 0.09 eV/atom, which suggests that F atoms are favorable to occupy octahedral sites on the surface.
This result is in perfect agreement with prior experiments where F termination is always observed at octahedral sites~\cite{schultz2019cm}. 

When the surface termination changes to O atoms occupying octahedral sites, a similar o-MXene is observed: Ti atoms segregate to the M' layers while Mo atoms remain inside M'' layers (Fig.~\ref{fig:Fig2}(e)).
However, a different chemical ordering emerges when O atoms terminate at the prismatic sites. In this case, Mo atoms segregate to M' layers and Ti atoms remain in the M'' layers, see Fig.~\ref{fig:Fig2}(f). 
This o-MXene configuration with Mo-rich M' and Ti-rich M'' sites is the most widely recognized chemical ordering in (TiMo)-based MXenes based on experimental observations~\cite{hong2020mrs}. This suggests that, in the experiments, O atoms possibly terminate at the prismatic sites. 
By performing more detailed structural analysis of the above o-MXenes, we find that the Ti-rich M' and Mo-rich M'' configurations in Fig.~\ref{fig:Fig2}(c-e) always exhibit larger interlayer distance \rev{($d$)} between M'' layers ($d_\textrm{M''-M''}$) than M'-M'' ones ($d_\textrm{M'-M''}$), but Mo-rich M' and Ti-rich M'' configuration (Fig.~\ref{fig:Fig2}(f)) shows nearly the same $d$ among the three M layers.
This demonstrates the effects of chemical ordering on local structures in DTM MXenes. 
The similar results obtained between our DFT/MC-simulated and fully-optimized MXene structures also highlight the efficiency of the improved DFT/MC code in capturing the local structural change induced by chemical ordering. 

\begin{figure*}[ht!]
\centering
\dimendef\prevdepth=0
{
\includegraphics[width=0.95\textwidth]{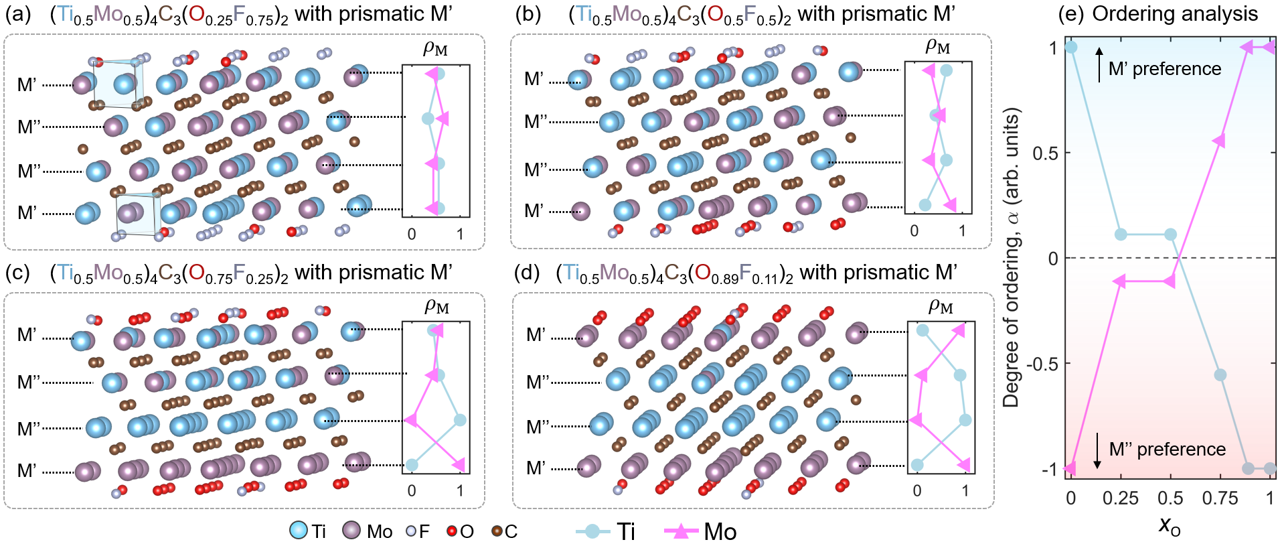}
}
\caption{Chemical ordering of (TiMo)-based DTM MXenes with varying O/F surface termination ratios.
%
DFT/MC-identified MECs of (\(\mathrm{Ti}_{0.5}\mathrm{Mo}_{0.5})_{4}\mathrm{C}_{3}\)(O\(_x\)F\(_{1-x}\))\(_2\) MXenes with increasing O content: (a) $x$ = 0.25, (b) $x$ = 0.50, (c) $x$ = 0.75, and (d) $x$ = 0.89.
The compositional profiles $\rho_M$ for both Ti and Mo atoms across the four M layers are shown for each case.
(e) Ordering analysis based on degree of ordering parameter, $\alpha$, as a function of O concentration on surface, $x_\textrm{O}$, for both Ti and Mo atoms.
The energy profiles $vs.$ MC iterations are provided in Fig. S2.
}
\label{fig:Fig3}
\end{figure*}

\begin{figure*}[ht!]
\centering
\dimendef\prevdepth=0
{
\includegraphics[width=1.0\textwidth]{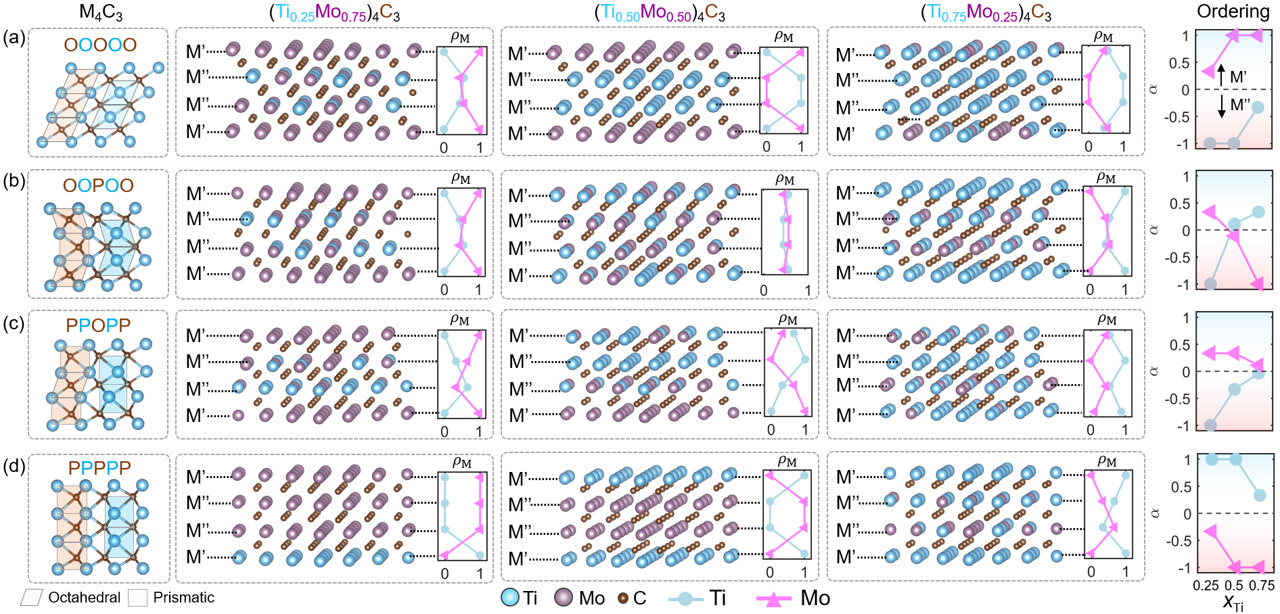}
}
\caption{Effects of atomic coordination in bare (TiMo)-based DTM MXenes on local chemical ordering.
Schematic of M$_4$C$_3$ MXenes with (a) OOOOO, (b) OOPOO, (c) PPOPP, and (d) PPPPP coordination and their DFT/MC-simulated MECs of three (TiMo)-based DTM MXenes, (Ti$_x$Mo$_{1-x}$)$_4$C$_3$, with $x$ = 0.25, 0.5, and 0.75.
The compositional profiles $\rho_M$ for both Ti and Mo atoms across the four M layers are shown for each case.
Degree of ordering parameter, $\alpha$, of both Ti and Mo atoms are plotted as a function of Ti content ($x_\textrm{Ti}$) for each coordination system.
The energy profiles $vs.$ MC iterations are provided in Fig. S3.
}
\label{fig:Fig4}
\end{figure*}

Furthermore, since different o-MXenes are observed when surface termination is switched from F to O termination at the prismatic sites (Fig.~\ref{fig:Fig2}d and f), we next examine how the ratio of F and O atoms influences the chemical ordering in these MXenes. 
Starting from the fully F-terminated case, we gradually increase the O concentration on the surfaces and performed DFT/MC simulations to search for their MECs. 
Figures~\ref{fig:Fig3}(a-d) illustrate the MECs of four (\(\mathrm{Ti}_{0.5}\mathrm{Mo}_{0.5})_{4}\mathrm{C}_{3}\)(O\(_x\)F\(_{1-x}\))\(_2\) MXenes with increasing O content: $x_\textrm{O}$ = 0.25, 0.50, 0.75, and 0.89. 
For $x_\textrm{O}$ = 0.25 and 0.50, DFT/MC-identified MECs show they exhibit highly disordered M layers, as no clear interlayer segregation of both Ti and Mo atoms is observed (Fig.~\ref{fig:Fig3}(a-b)).
These disordered structures are further supported by the 2D compositional profiles ($\rho_\textrm{M}$, M = Ti or Mo), which show nearly flat distributions of both Ti amd Mo atoms across all four M layers (Fig.~\ref{fig:Fig3}(a-b)).
When $x_\textrm{O}$ increases to 0.75, Fig.~\ref{fig:Fig3}(c) reveals a partial ordering, where Mo and Ti atoms enrich the bottom M' and M'' layers, respectively, while the upper M' and M'' layers remain a solid-solution-like (disordered) structure.
As $x_\textrm{O}$ further increases to 0.89, a clear o-MXene forms: Mo atoms segregate strongly to both outer M' layers and Ti atoms preferentially occupy the two inner M'' layers.

To quantify the order-to-disorder transition with different F/O ratios, the degree of ordering parameter ($\alpha$) for each MXene shown in Fig.~\ref{fig:Fig3} is calculated using $\alpha$ = ($\rho_\textrm{M'}$-$\rho_\textrm{M''}$)/($\rho_\textrm{M'}$+$\rho_\textrm{M''}$)~\cite{wyatt2025science}, where $\rho_\textrm{M'}$ and $\rho_\textrm{M''}$ are the composition of metal M (M = Ti or Mo) at the M' and M'' layers, respectively. 
In this definition, a positive value with $\alpha$ = 1 represents a preference for the metal to segregate to the M' layer; a negative value with $\alpha$ = -1 means a preference to segregate to the M'' layers; and $\alpha$ = 0 suggests no preferential segregation and thus indicate a disordered structure.
Fig.~\ref{fig:Fig3}(e) illustrates that as $x_\textrm{O}$ increases from 0 to 1, the (TiMo)-based DTM MXenes first undergo a transition from an ordered Ti-rich M' and Mo-rich M'' phase to a disordered phase as $x_\textrm{O}$ increases from 0 to 0.50, and then this disordered structure transforms back to another ordered phase with Mo-rich M' and Ti-rich M'' layers as $x_\textrm{O}$ continues to increase from 0.5 to 1.  
%

With a clear understanding of how surface termination elements and their coordination affect order-to-disorder transition in (TiMo)-based DTM MXenes, we next investigate how different atomic coordination of the MXene lattice and the composition of transition metals influence this transition behavior.
Four types of bare MXenes were constructed based on our prior work~\cite{oyeniran2025afm}, including ($i$) a purely octahedral MXene with OOOOO coordination, ($ii$) a twinned MXene with OOPOO coordination, ($iii$) a MXene with PPOPP coordination, and ($iv$) a fully prismatic MXene with PPPPP coordination, see Fig.~\ref{fig:Fig4}.
For each configuration, three compositions were considered: (Ti\(_{0.25}\)Mo\(_{0.75}\))\(_4\)C\(_3\), (Ti\(_{0.50}\)Mo\(_{0.50}\))\(_4\)C\(_3\), and (Ti\(_{0.75}\)Mo\(_{0.25}\))\(_4\)C\(_3\), and DFT/MC simulations were subsequently performed to determine their MECs.

Figure~\ref{fig:Fig4} shows the simulated MECs for all (Ti$_x$Mo$_{1-x}$)$_4$C$_3$ MXenes across various coordination types and compositions ($x$ = 0.25, 0.5, and 0.75).  
For purely octahedral coordination (Fig.~\ref{fig:Fig4}(a)), o-MXenes with  Ti (Mo) segregation to M'' (M') layers are always formed regardless of Ti concentration ($x_\textrm{Ti}$).
This can be attributed to favorable coordination environment for Ti-C bonding, which stabilizes Ti in the inner M'' layers and consequently pushes Mo to the outer M' layers~\cite{oyeniran2025afm}.
The ordered structure in purely octahedral coordination is also supported by the ordering analysis based on $\alpha$ presented in the right column of Fig.~\ref{fig:Fig4}(a).

In the twinned MXene with OOPOO coordination (Fig.~\ref{fig:Fig4}(b)), a similar o-MXene configuration is observed for $x_\textrm{Ti}$ = 0.25, where Mo atoms segregate to outer M' layers and Ti atoms remain in the inner two M'' layers.
Yet, as $x_\textrm{Ti}$ increases to 0.5, the MXene structure becomes highly disordered, but then transitions back to a more ordered structure at $x_\textrm{Ti}$ = 0.75, forming Ti-rich M' layers and Mo-rich M'' layers (Fig.~\ref{fig:Fig4}(b)).
Notably, despite only a slight change in the coordination between the OOPOO and OOOOO configurations, the equimolar (Ti\(_{0.5}\)Mo\(_{0.5}\))\(_4\)C\(_3\) MXenes exhibit completely different structures: the OOOOO configuration has a chemically ordered structure ($\alpha_\textrm{Ti}$ = -1 and $\alpha_\textrm{Mo}$ = 1), but OOPOO is highly disordered ($\alpha_\textrm{Ti}$ = 0.11 and $\alpha_\textrm{Mo}$ = -0.11), see Fig.~\ref{fig:Fig4}(a-b).
This difference demonstrates the critical role of local atomic coordination in governing chemical ordering and associated transition behaviors in DTM MXenes.   

As prismatic coordination becomes more dominant in DTM MXenes, the PPOPP and PPPPP structures show more interesting and complex ordering patterns and transition behaviors, as illustrated in Fig.~\ref{fig:Fig4}(c-d).
Specifically, in the PPOPP configuration, the MXene with $x_\textrm{Ti}$ = 0.25 shows the Mo segregation to the outer M' layers and Ti remaining in the M'' layers (Fig.~\ref{fig:Fig4}c), which is similar to the chemical ordering observed in the OOOOO and OOPOO configurations with same composition (Fig.~\ref{fig:Fig4}a-b). 
However, unlike the symmetric ordering observed in the octahedral-dominated MXenes, the two Ti-rich M'' layers in the PPOPP structure exhibit different $\rho_\textrm{Ti}$ values, leading to an asymmetric ordering pattern (Fig.~\ref{fig:Fig4}c).
As $x_\textrm{Ti}$ increases to 0.5, a new asymmetric ordering pattern appears, where Ti atoms now preferentially occupy the top M' and M'' layers, while Mo atoms segregate to the bottom M' and M'' layers (Fig.~\ref{fig:Fig4}c). 
Finally, at $x_\textrm{Ti}$ = 0.75, the MXene structure exhibits a complex, alternating disorder-order patterns, featuring a disordered top M', an ordered Ti-rich top M'', a disordered bottom M'', and an ordered Ti-rich bottom M' layer (Fig.~\ref{fig:Fig4}(c)).  

When the MXene is in a completely prismatic coordination (PPPPP, Fig.~\ref{fig:Fig4}(d)), the MXene with $x_\textrm{Ti}$ = 0.25 has an asymmetric chemical ordering pattern, where all Ti atoms segregate to the bottom M' layer, while the other three M layers are fully occupied by Mo atoms.
Increasing $x_\textrm{Ti}$ to 0.5 results in a clear o-MXene ($\alpha_\textrm{Ti}$ = 1 and $\alpha_\textrm{Mo}$ = -1), characterized by the Ti-rich M' layers and Mo-rich M'' layers (Fig.~\ref{fig:Fig4}(d)). 
This equimolar o-MXene has a completely reversed chemical ordering compared to the purely octahedral case in Fig.~\ref{fig:Fig4}(a). This reversal can be explained by the energetically favorable Mo-C bonding in prismatic coordination, which stabilizes Mo in the inner M'' layers and drives Ti atoms to segregate to the outer layers.
As $x_\textrm{Ti}$ further increases to 0.75, the MXene again shows asymmetric chemical ordering, as the two inner M'' layers contain unequal Mo concentrations, see Fig.~\ref{fig:Fig4}(d). 

\rev{In conclusion, using an improved DFT/MC framework, we reveal previously reported order-to-disorder transitions in (TiMo)-based DTM MXenes driven by changes in surface termination and/or local coordination environments.}
Our findings reveal that the formation of o-MXene with Mo-rich M'and Ti-rich M'' layers is only driven by the O termination at prismatic sites.
Conversely, O termination at octahedral sites and F termination at both prismatic and octahedral sites promote the formation of o-MXenes with reversed chemical ordering. 
Furthermore, order-to-disorder transitions of DTM MXenes can be driven by changing the F/O ratio on surface termination or by alternating the atomic coordination within the MXene lattices.
Although these findings are based on DTM MXenes, we anticipate that they are also be applicable to other multicomponent MXenes. 

\rev{Despite these promising findings, several important points should be noted. 
First, spin polarization is not included in the DFT/MC screening of (TiMo)-based DTM MXenes due to their small or no magnetization. However, other DTM MXenes, such as Ti$_2$WC$_2$T$_x$ and Cr$_2$TiC$_2$T$_x$, have been reported to exhibit magnetization~\cite{sakhraoui2025acs, hantan2020nh}.
This suggests that future DFT/MC screening of these MXenes should consider spin polarization, as local magnetization could potentially influence the chemical ordering in these magnetic MXenes. 
%
Second, all DFT/MC searches in this work are performed at 0 K, where thermal effects are negligible.
At elevated temperatures, MXenes may undergo structural transformation among different metastable phases~\cite{oyeniran2026}.
Because these metastable MXenes exhibit distinct local coordination environments, it is anticipated that the order-to-disorder transitions may occur concomitantly with structural transformation among these metastable phases. 
%
}

%
%
%
%
%
%

More importantly, the \rev{improved} DFT/MC framework is not limited to simulate the chemical ordering in DTM MXenes, but can be easily extended to any multicomponent MXenes.
In addition, the DFT/MC code is highly versatile. It can be applied to investigate a wide range of physical phenomena, including short-range ordering~\cite{antillon2020acta, liu2021cms}, interfacial segregation~\cite{hu2024cms}, and many other complex behaviors in multicomponent systems. 
By making the DFT/MC code publicly accessible, this framework is anticipated to achieve broader applicability and significant impacts on many research areas within materials science and nanoscience. 

\par{

\section*{Declaration of competing interest}
\par{
The authors declare that they have no known competing financial interests or personal relationships that could have appeared to influence the work reported. 
}

\section*{Data availability}
\par{
\rev{The improved DFT/MC code is available at https://github.com/huhuhhhh/DFT-MC, and additional data are available upon request from the corresponding author.}
}

\section*{Acknowledgements}
\par{
N.O. and C.H. acknowledge the support of U.S. Department of Energy (DOE) Office of Science, Basic Energy Sciences under Award No. DE-SC0025431.
This research used resources of the National Energy Research Scientific Computing Center, a DOE Office of Science User Facility supported by the Office of Science of the U.S. Department of Energy under Contract No. DE-AC02-05CH11231 using NERSC award BES-ERCAP0031213.
This work was also supported by a user project at the Center for Nanophase Materials Sciences (CNMS), a US DOE Office of Science User Facility, operated at Oak Ridge National Laboratory. Computations used resources of the National Energy Research Scientific Computing Center (NERSC), a US DOE Office of Science User Facility using NERSC award BES-ERCAP0027465 and ERCAP0031261 and BES-ERCAP35988.
}
\clearpage

\bibliography{jobname}

@article{naguib2011am,
  title={Two-dimensional nanocrystals produced by exfoliation of {T}i$_3${A}l{C}$_2$},
  author={Naguib, Michael and Kurtoglu, Murat and Presser, Volker and Lu, Jun and Niu, Junjie and Heon, Min and Hultman, Lars and Gogotsi, Yury and Barsoum, Michel W},
  journal={Adv. Mater.},
  volume={23},
  issue={37},
  pages={4248--4253},
  year={2011},
  publisher={Wiley Publishing},
  doi={10.1002/adma.201102306}
}

@article{naguib2012two,
  title={Two-dimensional transition metal carbides},
  author={Naguib, Michael and Mashtalir, Olha and Carle, Joshua and Presser, Volker and Lu, Jun and Hultman, Lars and Gogotsi, Yury and Barsoum, Michel W},
  journal={ACS Nano},
  volume={6},
  number={2},
  pages={1322--1331},
  year={2012},
  publisher={ACS Publications},
  doi={10.1021/nn204153h}
}

@article{gogotsi2023future,
  title={The future of {MX}enes},
  author={Gogotsi, Yury},
  journal={Chem. Mater.},
  volume={35},
  number={21},
  pages={8767--8770},
  year={2023},
  publisher={ACS Publications},
  doi={10.1021/acs.chemmater.3c02491}
}

@article{li2022review,
  title={{MX}ene chemistry, electrochemistry and energy storage applications},
  author={Li, Xinliang and Huang, Zhaodong and Shuck, Christopher E and Liang, Guojin and Gogotsi, Yury and Zhi, Chunyi},
  journal={Nat. Rev. Chem.},
  volume={6},
  number={6},
  pages={389--404},
  year={2022},
  publisher={Nature Publishing Group UK London},
  doi={10.1038/s41570-022-00384-8}
}

@article{lim2022review,
  title={Fundamentals of {MX}ene synthesis},
  author={Lim, Kang Rui Garrick and Shekhirev, Mikhail and Wyatt, Brian C and Anasori, Babak and Gogotsi, Yury and Seh, Zhi Wei},
  journal={Nat. Synth.},
  volume={1},
  number={8},
  pages={601--614},
  year={2022},
  publisher={Nature Publishing Group UK London},
  doi={10.1038/s44160-022-00104-6}
}

@article{anasori2017review,
  title={2{D} metal carbides and nitrides ({MX}enes) for energy storage},
  author={Anasori, Babak and Lukatskaya, Maria R and Gogotsi, Yury},
  volume={2},
  number={16098},
  journal = {Nat. Rev. Mater.},
  pages={677--722},
  year={2017},
  publisher={Nature publishing},
  doi={10.1038/natrevmats.2016.98}
}

@article{kamy2020science,
  title={Covalent surface modifications and superconductivity of two-dimensional metal carbide {MX}enes},
  author={Kamysbayev, Vladislav and Filatov, Alexander S and Hu, Huicheng and Rui, Xue and Lagunas, Francisco and Wang, Di and Klie, Robert F and Talapin, Dmitri V},
  journal={Science},
  volume={369},
  number={6506},
  pages={979--983},
  year={2020},
  publisher={American Association for the Advancement of Science},
  doi={10.1126/science.aba8311}
}

@article{oyeniran2025apl,
  title={First-principles and {M}onte {C}arlo simulations of high-entropy {MX}enes},
  author={Oyeniran, Noah and Chowdhury, Oyshee and Hu, Chongze},
  journal={Appl. Phys. Lett.},
  volume={126},
  number={12},
  year={2025},
  publisher={AIP Publishing},
  doi={10.1063/5.0258487}
}

@article{kresse1993ab,
  title={Ab initio molecular dynamics for liquid metals},
  author={Kresse, Georg and Hafner, J{\"u}rgen},
  journal={Phys. Rev. B},
  volume={47},
  number={1},
  pages={558},
  year={1993},
  publisher={APS},
  doi={10.1103/PhysRevB.47.558}
}

@article{kresse1996efficiency,
  title={Efficiency of ab-initio total energy calculations for metals and semiconductors using a plane-wave basis set},
  author={Kresse, Georg and Furthm{\"u}ller, J{\"u}rgen},
  journal={Comput. Mater. Sci.},
  volume={6},
  number={1},
  pages={15--50},
  year={1996},
  publisher={Elsevier},
  doi={10.1016/0927-0256(96)00008-0}
}

@article{kresse1996efficient,
  title={Efficient iterative schemes for ab initio total-energy calculations using a plane-wave basis set},
  author={Kresse, Georg and Furthm{\"u}ller, J{\"u}rgen},
  journal={Phys. Rev. B},
  volume={54},
  number={16},
  pages={11169},
  year={1996},
  publisher={APS},
  doi={10.1103/PhysRevB.54.11169}
}

@article{metropolis1953equation,
  title={Equation of state calculations by fast computing machines},
  author={Metropolis, Nicholas and Rosenbluth, Arianna W and Rosenbluth, Marshall N and Teller, Augusta H and Teller, Edward},
  journal={J. Chem. Phys.},
  volume={21},
  number={6},
  pages={1087--1092},
  year={1953},
  publisher={American Institute of Physics},
  doi={10.1063/1.1699114}
}

@article{blochl1994projector,
  title={Projector augmented-wave method},
  author={Bl{\"o}chl, Peter E},
  journal={Phys. Rev. B},
  volume={50},
  number={24},
  pages={17953},
  year={1994},
  publisher={APS},
  doi={10.1103/PhysRevB.50.17953}
}

@article{kresse1999ultrasoft,
  title={From ultrasoft pseudopotentials to the projector augmented-wave method},
  author={Kresse, Georg and Joubert, Daniel},
  journal={Phys. Rev. B},
  volume={59},
  number={3},
  pages={1758},
  year={1999},
  publisher={APS},
  doi={10.1103/PhysRevB.59.1758}
}

@article{perdew1996generalized,
  title={Generalized gradient approximation made simple},
  author={Perdew, John P and Burke, Kieron and Ernzerhof, Matthias},
  journal={Phys. Rev. Lett.},
  volume={77},
  number={18},
  pages={3865},
  year={1996},
  publisher={APS},
  doi={10.1103/PhysRevLett.77.3865}
}

@article{grimme2006semiempirical,
  title={Semiempirical GGA-type density functional constructed with a long-range dispersion correction},
  author={Grimme, Stefan},
  journal={Comput. Chem},
  volume={27},
  number={15},
  pages={1787--1799},
  year={2006},
  publisher={Wiley Online Library},
  doi={10.1002/jcc.20495}
}

@article{oyeniran2025afm,
  title={A panoramic view of {MX}enes via an atomic coordination-based design strategy},
  author={Oyeniran, Noah and Chowdhury, Oyshee and Hu, Chongze and Dumitrica, Traian and Ganesh, Panchapakesan and Jakowski, Jacek and Chen, Zhongfang and Unocic, Raymond R and Naguib, Michael and Meunier, Vincent and others},
  journal={Adv. Funct. Mater.},
  pages={e08047},
  year={2025},
  publisher={Wiley Online Library},
   doi={10.1002/adfm.202508047}
}

@article{oyeniran2026,
  title={First-Principles Investigation of Structure-Property Relationships in Stable and Metastable MXenes},
  author={Oyeniran, Noah and Das, Shimanta and Dumitrica, Traian and Ganesh, Panchapakesan and Jakowski, Jacek and Chen, Zhongfang and Unocic, Raymond R and Naguib, Michael and Meunier, Vincent and others},
  journal={In preparation},
  pages={},
  year={2026},
  publisher={},
  doi={}
}

@article{hanan2025comprehensive,
  title={A comprehensive review of double transition metal {MX}ene (Mo$_2$Ti$_2$C$_3$T$_x$) in energy storage, Conversion, and Harvesting},
  author={Hanan, Abdul and Bibi, Faiza and Elsa, George and Numan, Arshid and Walvekar, Rashmi and Khalid, Mohammad},
  journal={Mater. Today Energy},
  pages={101905},
  year={2025},
  publisher={Elsevier},
  doi={10.1016/j.mtener.2025.101905}
}

@article{hussain2024double,
  title={Double transition-metal MXenes: Classification, properties, machine learning, artificial intelligence, and energy storage applications},
  author={Hussain, Iftikhar and Sajjad, Uzair and Kewate, Onkar Jaywant and Amara, Umay and Bibi, Faiza and Hanan, Abdul and Potphode, Darshna and Ahmad, Muhammad and Javed, Muhammad Sufyan and Rosaiah, P and others},
  journal={Mater. Today Phys},
  volume={42},
  pages={101382},
  year={2024},
  publisher={Elsevier},
  doi={10.1016/j.mtphys.2024.101382}
}

@article{chen2025high,
  title={High-entropy {MX}ene-based multiple soliton states generation},
  author={Chen, Yimiao and Zheng, Wenchen and Zhuo, Yijian and Zhang, Yule and Wei, Songrui and Du, Bowen and Chen, Weichen and Jiang, Ke and Ge, Yanqi and others},
  journal={J. Light. Technol.},
  year={2025},
  publisher={IEEE},
  doi={10.1109/JLT.2025.3554518}
}

@article{das2023perspective,
  title={Perspective on high entropy {MX}enes for energy storage and catalysis},
  author={Das, Pratteek and Dong, Yanfeng and Wu, Xianhong and Zhu, Yuanyuan and Wu, Zhong-Shuai},
  journal={Sci bull},
  volume={68},
  number={16},
  pages={1735--1739},
  year={2023},
  doi={10.1016/j.scib.2023.07.022}
}

@article{leong2022elucidating,
  title={Elucidating the chemical order and disorder in high-entropy {MX}enes: A high-throughput survey of the atomic configurations in TiVNbMoC$_3$ and TiVCrMoC$_3$},
  author={Leong, Zhidong and Jin, Hongmei and Wong, Zicong Marvin and Nemani, Kartik and Anasori, Babak and Tan, Teck Leong},
  journal={Chem. Mater.},
  volume={34},
  number={20},
  pages={9062--9071},
  year={2022},
  publisher={ACS Publications},
  doi={10.1021/acs.chemmater.2c01673}
}

@article{wyatt20212d,
  title={2{D} {MX}enes: tunable mechanical and tribological properties},
  author={Wyatt, Brian C and Rosenkranz, Andreas and Anasori, Babak},
  journal={Adv. Mater.},
  volume={33},
  number={17},
  pages={2007973},
  year={2021},
  publisher={Wiley Online Library},
  doi={10.1002/adma.202007973}
}

@article{zhang2025ultra,
  title={Ultra-rapid synthesis of high-entropy MAX phases and their derivative {MX}enes for battery electrodes},
  author={Zhang, Liang and Li, Huicong and Zhang, Xiaoyu and Liu, Chunxue and Sun, Yifei and Zhang, Yiyuan and Fang, Zhen and He, Jiangang and Wang, Rongming and Jiang, Kai and others},
  journal={Angew. Chem.-Int. Ed.},
  volume={137},
  number={6},
  pages={e202418538},
  year={2025},
  publisher={Wiley Online Library},
  doi={10.1002/anie.202418538}
}

@article{biswas2022mxene,
  title={{MX}ene: evolutions in chemical synthesis and recent advances in applications},
  author={Biswas, Sayani and Alegaonkar, Prashant S},
  journal={Surfaces},
  volume={5},
  number={1},
  pages={1--34},
  year={2022},
  publisher={Multidisciplinary Digital Publishing Institute},
  doi={10.3390/surfaces5010001}
}

@article{nemani2021high,
  title={High-entropy 2D carbide mxenes: TiVNbMoC$_3$ and TiVCrMoC$_3$},
  author={Nemani, Srinivasa Kartik and Zhang, Bowen and Wyatt, Brian C and Hood, Zachary D and Manna, Sukriti and Khaledialidusti, Rasoul and Hong, Weichen and Sternberg, Michael G and Sankaranarayanan, Subramanian KRS and Anasori, Babak},
  journal={ACS nano},
  volume={15},
  number={8},
  pages={12815--12825},
  year={2021},
  publisher={ACS Publications},
  doi={10.1021/acsnano.1c02775}
}

@article{wyatt2025science,
  title={Order-to-disorder transition due to entropy in layered and 2{D} carbides},
  author={Wyatt, Brian C and Yang, Yinan and Micha{\l}owski, Pawe{\l} P and Parker, Tetiana and Morency, Yamil{\'e}e and Urban, Francesca and Kadagishvili, Givi and Tanwar, Manushree and Muhoza, Sixbert P and Nemani, Srinivasa Kartik and others},
  journal={Science},
  volume={389},
  number={6764},
  pages={1054--1058},
  year={2025},
  publisher={American Association for the Advancement of Science},
  doi={10.1126/science.adv4415}
}

@article{2021science,
  title={The world of two-dimensional carbides and nitrides ({MX}enes)},
  author={VahidMohammadi, Armin and Rosen, Johanna and Gogotsi, Yury},
  journal={Science},
  volume={372},
  number={6547},
  pages={eabf1581},
  year={2021},
  doi={10.1126/science.abf1581}
}

@article{khan2023review,
  title={Recent advances in {MX}enes: a future of nanotechnologies},
  author={Khan, Karim and Tareen, Ayesha Khan and Iqbal, Muhammad and Hussain, Iftikhar and Mahmood, Asif and Khan, Usman and Khan, Muhammad Farooq and Zhang, Han and Xie, Zhongjian},
  journal={J. Mater. Chem. A},
  volume={11},
  number={37},
  pages={19764--19811},
  year={2023},
  publisher={Royal Society of Chemistry},
  doi={10.1039/D3TA03069E}
}

@article{rems2024cm,
  title={Pivotal role of surface terminations in {MX}ene thermodynamic stability},
  author={Rems, Ervin and Hu, Yong-Jie and Gogotsi, Yury and Dominko, Robert},
  journal={Chem. Mater.},
  volume={36},
  number={20},
  pages={10295--10306},
  year={2024},
  publisher={ACS Publications},
  doi={10.1021/acs.chemmater.4c02274}
}

@article{george2019natrevmater,
  title={High-entropy alloys},
  author={George, Easo P and Raabe, Dierk and Ritchie, Robert O},
  journal={Nat. Rev. Mater.},
  volume={4},
  number={8},
  pages={515--534},
  year={2019},
  doi={10.1038/s41578-019-0121-4}
}

@article{oses2020natrevmater,
  title={High-entropy ceramics},
  author={Oses, Corey and Toher, Cormac and Curtarolo, Stefano},
  journal={Nat. Rev. Mater.},
  volume={5},
  number={4},
  pages={295--309},
  year={2020},
  doi={10.1038/s41578-019-0170-8}
}

@article{han2024natrevmater,
  title={Multifunctional high-entropy materials},
  author={Han, Liuliu and Zhu, Shuya and Rao, Ziyuan and Scheu, Christina and Ponge, Dirk and Ludwig, Alfred and Zhang, Hongbin and Gutfleisch, Oliver and Hahn, Horst and Li, Zhiming and others},
  journal={Nat. Rev. Mater.},
  pages={1--20},
  year={2024},
  doi={10.1038/s41578-024-00720-y}
}

@article{miracle2017acta,
  title={A critical review of high entropy alloys and related concepts},
  author={Miracle, Daniel B and Senkov, Oleg N},
  journal={Acta Mater.},
  volume={122},
  pages={448--511},
  year={2017},
  doi={10.1016/j.actamat.2016.08.081}
}

@article{qin2020jecs,
  title={Dual-phase high-entropy ultra-high temperature ceramics},
  author={Qin, Mingde and Gild, Joshua and Hu, Chongze and Wang, Haoren and Hoque, Md Shafkat Bin and Braun, Jeffrey L and Harrington, Tyler J and Hopkins, Patrick E and Vecchio, Kenneth S and Luo, Jian},
  journal={J. Europ. Ceram. Soc.},
  volume={40},
  number={15},
  pages={5037--5050},
  year={2020},
  doi={10.1016/j.jeurceramsoc.2020.05.040}
}

@article{qin2020scripta,
  title={High-entropy monoborides: Towards superhard materials},
  author={Qin, Mingde and Yan, Qizhang and Wang, Haoren and Hu, Chongze and Vecchio, Kenneth S and Luo, Jian},
  journal={Scr. Mater.},
  volume={189},
  pages={101--105},
  year={2020},
  doi={10.1016/j.scriptamat.2020.08.018}
}

@article{hong2020mrs,
  title={Double transition-metal {MX}enes: Atomistic design of two-dimensional carbides and nitrides},
  author={Hong, Weichen and Wyatt, Brian C and Nemani, Srinivasa Kartik and Anasori, Babak},
  journal={MRS Bull.},
  volume={45},
  number={10},
  pages={850--861},
  year={2020},
  publisher={Cambridge University Press},
  doi={10.1557/mrs.2020.251}
}

@article{nemani2023commat,
  title={Functional two-dimensional high-entropy materials},
  author={Nemani, Srinivasa Kartik and Torkamanzadeh, Mohammad and Wyatt, Brian C and Presser, Volker and Anasori, Babak},
  journal={Commun. Mater.},
  volume={4},
  number={1},
  pages={16},
  year={2023},
  doi={10.1038/s43246-023-00341-y}
}

@article{meshkian2017acta,
  title={Theoretical stability and materials synthesis of a chemically ordered MAX phase, Mo$_2$ScAlC$_2$, and its two-dimensional derivate Mo$_2$ScC$_2$ MXene},
  author={Meshkian, Rahele and Tao, Quanzheng and Dahlqvist, Martin and Lu, Jun and Hultman, Lars and Rosen, Johanna},
  journal={Acta Mater.},
  volume={125},
  pages={476--480},
  year={2017},
  publisher={Elsevier},
  doi={10.1016/j.actamat.2016.12.008}
}

@article{anasori2015acs,
  title={Two-dimensional, ordered, double transition metals carbides ({MX}enes)},
  author={Anasori, Babak and Xie, Yu and Beidaghi, Majid and Lu, Jun and Hosler, Brian C and Hultman, Lars and Kent, Paul RC and Gogotsi, Yury and Barsoum, Michel W},
  journal={ACS nano},
  volume={9},
  number={10},
  pages={9507--9516},
  year={2015},
  publisher={ACS Publications},
  doi={10.1021/acsnano.5b03591}
}

@article{maldonado2022acs,
  title={Atomic-scale understanding of Li storage processes in the Ti$_4$C$_3$ and chemically ordered Ti$_2$Ta$_2$C$_3$ MXenes: a theoretical and experimental assessment},
  author={Maldonado-Lopez, Daniel and Rodriguez, Jassiel R and Pol, Vilas G and Syamsai, Ravuri and Andrews, Nirmala Grace and Guti{\'e}rrez-Ojeda, SJ and Ponce-Perez, Rodrigo and Moreno-Armenta, Ma Guadalupe and Guerrero-Sanchez, Jonathan},
  journal={ACS Appl. Energy Mater.},
  volume={5},
  number={2},
  pages={1801--1809},
  year={2022},
  publisher={ACS Publications},
  doi={10.1021/acsaem.1c03239}
}

@article{thangavelu2025review,
  title={A review on {MX}ene terminations},
  author={Thangavelu, Hari Hara Sudhan and Huang, Changjie and Chabanais, Florian and Palisaitis, Justinas and Persson, Per O{\AA}},
  journal={Adv. Funct. Mater.},
  pages={e15604},
  year={2025},
  publisher={Wiley Online Library},
  doi={10.1002/adfm.202515604}
}

@article{zhou2023atomic,
  title={Atomic scale design of {MX}enes and their parent materials─ from theoretical and experimental perspectives},
  author={Zhou, Jie and Dahlqvist, Martin and Bjork, Jonas and Rosen, Johanna},
  journal={Chemical reviews},
  volume={123},
  number={23},
  pages={13291--13322},
  year={2023},
  publisher={ACS Publications},
  doi={10.1021/acs.chemrev.3c00241}
}

@article{hu2018acs,
  title={Ab initio predictions of strong interfaces in transition-metal carbides and nitrides for superhard nanocomposite coating applications},
  author={Hu, Chongze and Huang, Jingsong and Sumpter, Bobby G and Meletis, Efstathios and Dumitrica, Traian},
  journal={ACS Appl. Nano Mater.},
  volume={1},
  number={5},
  pages={2029--2035},
  year={2018},
  publisher={ACS Publications},
  doi={10.1021/acsanm.8b00227}
}

@article{hu2018jpcc,
  title={Screening surface structure of {MX}enes by high-throughput computation and vibrational spectroscopic confirmation},
  author={Hu, Tao and Hu, Minmin and Gao, Bo and Li, Wu and Wang, Xiaohui},
  journal={J. Phys. Chem. C},
  volume={122},
  number={32},
  pages={18501--18509},
  year={2018},
  publisher={ACS Publications},
  doi={10.1021/acs.jpcc.8b04427}
}

@article{schultz2019cm,
  title={Surface termination dependent work function and electronic properties of Ti$_3$C$_2$T$_x$ {MX}ene},
  author={Schultz, Thorsten and Frey, Nathan C and Hantanasirisakul, Kanit and Park, Soohyung and May, Steven J and Shenoy, Vivek B and Gogotsi, Yury and Koch, Norbert},
  journal={Chem. Mater.},
  volume={31},
  number={17},
  pages={6590--6597},
  year={2019},
  publisher={ACS Publications},
  doi={10.1021/acs.chemmater.9b00414}
}

@article{hu2024cms,
  title={Computational modeling of grain boundary segregation: A review},
  author={Hu, Chongze and Dingreville, R{\'e}mi and Boyce, Brad L},
  journal={Comput. Mater. Sci.},
  volume={232},
  pages={112596},
  year={2024},
  doi={10.1016/j.commatsci.2023.112596}
}

@article{antillon2020acta,
  title={Chemical short range order strengthening in a model FCC high entropy alloy},
  author={Antillon, E and Woodward, C and Rao, SI and Akdim, B and Parthasarathy, TA},
  journal={Acta Mater.},
  volume={190},
  pages={29--42},
  year={2020},
  publisher={Elsevier},
  doi={10.1016/j.actamat.2020.02.041}
}

@article{liu2021cms,
  title={Monte Carlo simulation of order-disorder transition in refractory high entropy alloys: A data-driven approach},
  author={Liu, Xianglin and Zhang, Jiaxin and Yin, Junqi and Bi, Sirui and Eisenbach, Markus and Wang, Yang},
  journal={Comput. Mater. Sci},
  volume={187},
  pages={110135},
  year={2021},
  publisher={Elsevier},
  doi={10.1016/j.commatsci.2020.110135}
}

@article{sakhraoui2025acs,
  title={Electronic and Magnetic Properties of Ordered Double Transition-Metal MXenes: Ti$_2$MC$_2$T$_2$ (M= Cr, Mo, and W; T= F, O, OH, and Cl)},
  author={Sakhraoui, Taoufik and Karlicky, Frantisek},
  journal={ACS omega},
  volume={10},
  number={45},
  pages={54061--54069},
  year={2025},
  publisher={ACS Publications}
}

@article{hantan2020nh,
  title={Evidence of a magnetic transition in atomically thin Cr$_2$TiC$_3$T$_x$ MXene},
  author={Hantanasirisakul, Kanit and Anasori, Babak and Nemsak, Slavomir and Hart, James L and Wu, Jiabin and Yang, Yizhou and Chopdekar, Rajesh V and Shafer, Padraic and May, Andrew F and Moon, Eun Ju and others},
  journal={Nanoscale Horiz.},
  volume={5},
  number={12},
  pages={1557--1565},
  year={2020},
  publisher={Royal Society of Chemistry}
}
\bibliographystyle{achemso}

\end{document}